\documentclass[twocolumn,showpacs,preprintnumbers,amsmath,amssymb,prl]{revtex4}
\usepackage{graphicx}
\usepackage{bm}
\usepackage{hyperref}
\begin{document}

\title{Collective behavior and critical fluctuations in the spatial
  spreading of obesity, diabetes and cancer}

\author{Lazaros K. Gallos$^1$, Pablo Barttfeld$^2$, Shlomo
  Havlin$^{3}$, Mariano Sigman$^2$, and Hern\'an A. Makse$^{1,2}$}

\affiliation{ $^1$ Levich Institute and Physics Department, City
  College of New York, New York, New York 10031, USA \\
  $^2$ Physics Department, FCEyN, Universidad de Buenos Aires, Buenos
  Aires, Argentina \\
  $^3$ Minerva Center and Physics Department, Bar-Ilan University,
  Ramat Gan 52900, Israel }

\date{\today}

\begin{abstract}

  Non-communicable diseases like diabetes, obesity and certain forms
  of cancer have been increasing in many countries at alarming
  levels. A difficulty in the conception of policies to reverse these
  trends is the identification of the drivers behind the global
  epidemics. Here, we implement a spatial spreading analysis to
  investigate whether non-communicable diseases like diabetes, obesity
  and cancer show spatial correlations revealing the effect of
  collective and global factors acting above individual
  choices. Specifically, we adapt a theoretical framework for critical
  physical systems displaying collective behavior to decipher the laws
  of spatial spreading of diseases. We find a regularity in the
  spatial fluctuations of their prevalence revealed by a pattern of
  scale-free long-range correlations.  The fluctuations are anomalous,
  deviating in a fundamental way from the weaker correlations found in
  the underlying population distribution. The resulting scaling
  exponents allow us to broadly classify the indicators into two
  universality classes, weakly or strongly correlated. This collective
  behavior indicates that the spreading dynamics of obesity, diabetes
  and some forms of cancer like lung cancer are analogous to a
  critical point of fluctuations, just as a physical system in a
  second-order phase transition. According to this notion, individual
  interactions and habits may have negligible influence in shaping the
  global patterns of spreading. Thus, obesity turns out to be a global
  problem where local details are of little importance.
  Interestingly, we find the same critical fluctuations in obesity and
  diabetes, and in the activities of economic sectors associated with
  food production such as supermarkets, food and beverage stores---
  which cluster in a different universality class than other generic
  sectors of the economy. These results motivate future interventions
  to investigate the causality of this relation providing guidance for
  the implementation of preventive health policies.


\end{abstract}
\maketitle

The World Health Organization has recognized obesity as a global
epidemic \cite{who}. Obesity heads the list of non-communicable
diseases (NCD) like diabetes and cancer, for which no prevention
strategy has managed to control their spreading
\cite{foresight,hill,swinburn,nestle,christakis,christakis2}. Here,
Since the gain of excessive body weight is related to an increase in
calories intake and physical inactivity \cite{popkin,cutler,haslam} a
principal aspect of prevention has been directed to individual habits
\cite{healthypeople}.
However, the prevalence of NCDs shows strong spatial clustering
\cite{schuurman,michimi,cdc}. Furthermore, obesity spreading has shown
high susceptibility to social pressure \cite{christakis} and global
economic drivers \cite{hill,swinburn,nestle,christakis2}. This
suggests that the spread and growth of obesity and other NCDs may be
governed by collective behavior acting over and above individual
factors such as genetics and personal choices
\cite{swinburn,nestle}. 

\begin{figure}
\centerline{\resizebox{10.0cm}{!} { \includegraphics{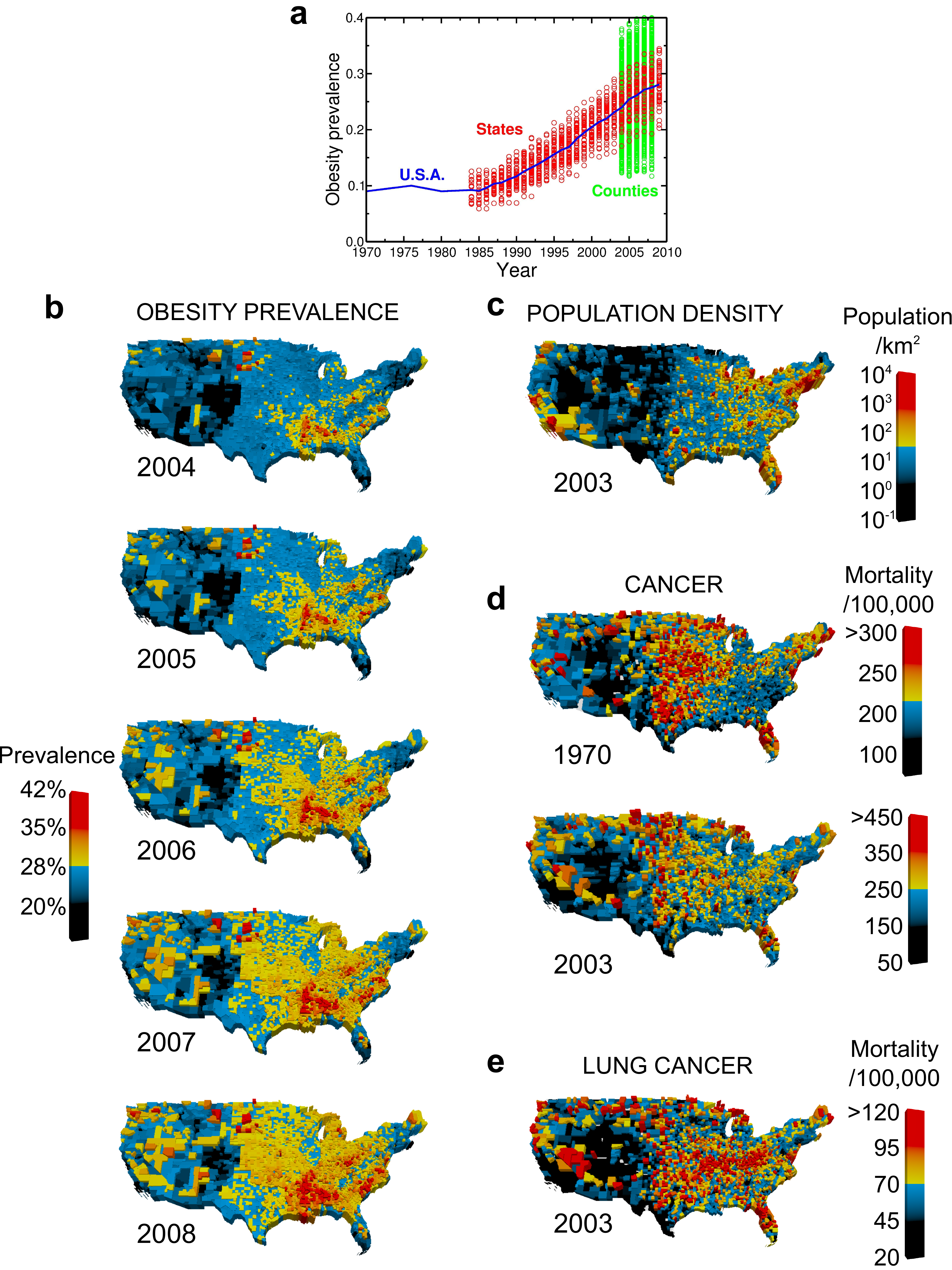}}}
\caption{{\bf The obesity transition.} {\bf a,} CDC
\cite{cdc} provides an estimate of the number of obese adults, based
on self-reported weight and height, country-wide since 1970 (blue
line), at the state level from 1984 to 2009 (red symbols), and at the
county level from 2004 to 2008 (green symbols).  A transition is
observed around 1980.  We base our analysis on the micro-data at the
county level.  {\bf b,} Map of the spatial spreading of obesity
prevalence evidencing clustering dynamics. {\bf c,} Map of the
population density defined at the county level in 2003 showing
correlated patterns albeit with less clustering than in obesity. {\bf
  d,} Map of cancer mortality rates per county in 1970 and 2003
visualizing the transition from high correlations and clustering to
weak correlation and more uniformity in 2003. {\bf e,} Map of lung
cancer mortality per county indicating large clustering properties
similar to obesity.}
\label{transition}
\end{figure}

To study the emergence of collective dynamics in the spatial spreading
of obesity and other NCDs, we implement a statistical clustering
analysis based on critical phenomenon physics.
We start by investigating regularities in obesity spreading derived
from correlation patterns of demographic variables. Obesity is
determined through the Body Mass Index (BMI) obtained via the formula
weight(kg)/[height (m)]$^2$.  The obesity prevalence is defined as the
percentage of adults aged $\ge$ 18 years with a BMI $\ge$ 30.  We
investigate the spatial correlations of obesity prevalence in the USA
during a specific year using micro-data defined at the county-level
provided by the US Centers for Disease Control (CDC) \cite{cdc}
through the Behavioral Risk Factor Surveillance System (BRFSS) from
2004 to 2008 (see Methods Section \ref{data}).
The average percentage of obesity in USA was historically around 10\%.
In the early 80s, an obesity transition in the hitherto robust
percentage, steeply increased the obesity prevalence
(Fig.~\ref{transition}a).

The spatial map of obesity prevalence in the USA shows that
neighboring areas tend to present similar percentages of obese
population \cite{cdc} forming spatial `obesity clusters'
\cite{schuurman,michimi}.  The evolution of the spatial map of obesity
from 2004 to 2008 at the county level (Fig.~\ref{transition}b)
highlights the mechanism of cluster growth. Characterizing such
geographical spreading presents a challenge to current theoretical
physics frameworks of cluster dynamics
\cite{stanley,shlomo,coniglio,weinrib1,sona,makse}.  The equal-time
two-point correlation function, $C(r)$, determines the properties of
such spatial arrangement by measuring the influence of an observable
$x_i$ in county $i$ (e.g., in this study: adult population density,
prevalence of obesity and diabetes, cancer mortality rates and
economic activity) on another county at distance $r$ \cite{stanley}:
\begin{equation}
  C(r) \equiv \frac{1}{\sigma_0}\frac{
  \sum_{ij} (x_i-\overline{x}) (x_j-\overline{x}) \delta(r_{ij}-r)}{
  \sum_{ij} \delta(r_{ij}-r)}.
\label{cr}
\end{equation}
Here, $\overline{x}$ is the average over all $N=3,141$ counties in the
USA, $\sigma_0=\sum_i(x_i-\overline{x})^2/N$ is the standard
deviation, $r_{ij}$ is the euclidean distance between the geometrical
center of counties $i$ and $j$.
Large positive values of $C(r)$ reveal strong correlations, while
negative values imply anti-correlations, i.e., two areas with opposed
tendencies relative to the mean in obesity prevalence (analogous to
two domains with opposite spins in a ferromagnet \cite{stanley}).

\begin{figure}
\centerline{\resizebox{6.0cm}{!} { \includegraphics{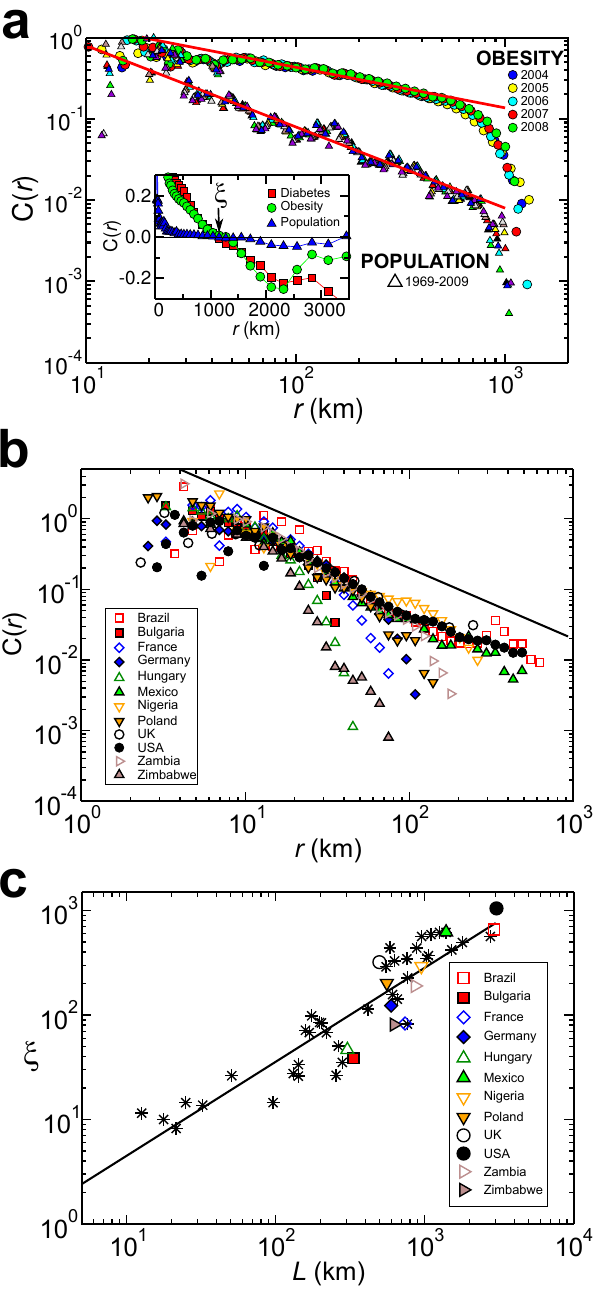}}}
\caption{{\bf Long-range correlations in spreading phenomena.} {\bf
    a,} Correlation function, $C(r)$, averaged over counties at
  distance $r$ for population density from 1969-2009 and obesity
  prevalence from 2004-2008.  The lines are fittings based on OLS
  regression analysis \cite{ols}. The inset shows the correlation
  length at $C(\xi)=0$ and highlights the fact that $\xi$ is
  approximately the same for the population, obesity and diabetes
  prevalence (data for 2004). The inset also highlights the
  anticorrelations for $r>\xi$.
%
{\bf b,} Population density correlation function, $C(r)$ vs $r$, for
different countries in 2009 as indicated.  {\bf c,} Correlation length
$\xi$ vs linear country size $L$ for different countries. The symbols
indicate the same countries as in Fig. \ref{correl}b. The remaining
star symbols are for other countries as indicated in Table
\ref{xi}. $L$ is the square root of the total area of the country.}
\label{correl}
\end{figure}

Spatial correlations in any indicator ought to be referred to the
natural correlations of population fluctuations
(Fig.~\ref{transition}c). To this aim, we first calculate $C(r)$ for
the population (adults $\ge$ 18 years) in USA counties, $p_i$, by
using the density: $x_i=p_i/a_i$ in Eq. (\ref{cr}), where $a_i$ is the
county area. Population density correlations show a slow fall-off with
distance (Fig. \ref{correl}a) described by a power-law up to a
correlation length $\xi$:
\begin{equation}
  C(r)\sim r^{-\gamma}, \,\,\, \,\, r\lesssim\xi,
  \label{gamma}
\end{equation}
where $\gamma$ is the correlation exponent.  Correlations become
short-ranged when $\gamma\ge d$ ($d=2$ is the dimension of the map),
and stronger as $\gamma$ decreases \cite{stanley,shlomo}. An Ordinary
Least Squares (OLS) regression analysis \cite{ols} on the population
reveals the exponent $\gamma=1.01\pm0.08$ (averaged over 1969-2009,
Fig.  ~\ref{correl}a, error bars denote 95\% confidence interval
[CI]). The inset of Fig. \ref{correl}a reveals a distance where
correlations vanish, $C(\xi)=0$ with $\xi=1050$km, representing the
average size of the correlated domains \cite{cavagna}. As we increase
$r$ larger than $\xi$, we consider correlations between areas in the
East and West which are anti-correlated since $C(r)<0$ for $r>\xi$.

To determine whether population correlations are scale-free, we
calculate $C(r)$ for geographical systems of different sizes using a
high resolution grid of 2.5 arc-seconds, available for several
countries from \cite{ciesin} (Methods Section \ref{gridded}).  The
resulting correlations (Fig. \ref{correl}b) reveal the same picture as
for the USA at the county-level (Fig. \ref{correl}a), i.e., a
power-law up to a correlation length.
We then measure $\xi$ for every country, and investigate whether, as
expected with the laws of critical phenomena \cite{stanley,mora}, it
increases with the country size, $L$. Indeed, we obtain
(Fig.~\ref{correl}c and  Table \ref{xi}),
\begin{equation}
  \xi(L) \sim L^{\nu},
\label{sf}
\end{equation}
where $\nu=0.9\pm0.1$ is the correlation length exponent
\cite{stanley}.
This result implies that the fluctuations in human agglomerations are
scale-free, i.e., the only length-scale in the system is set by its
size and the correlation length become infinite when $L\to \infty$
\cite{stanley,mora,cavagna}.

\begin{figure}
\centerline{\resizebox{9.0cm}{!} { \includegraphics{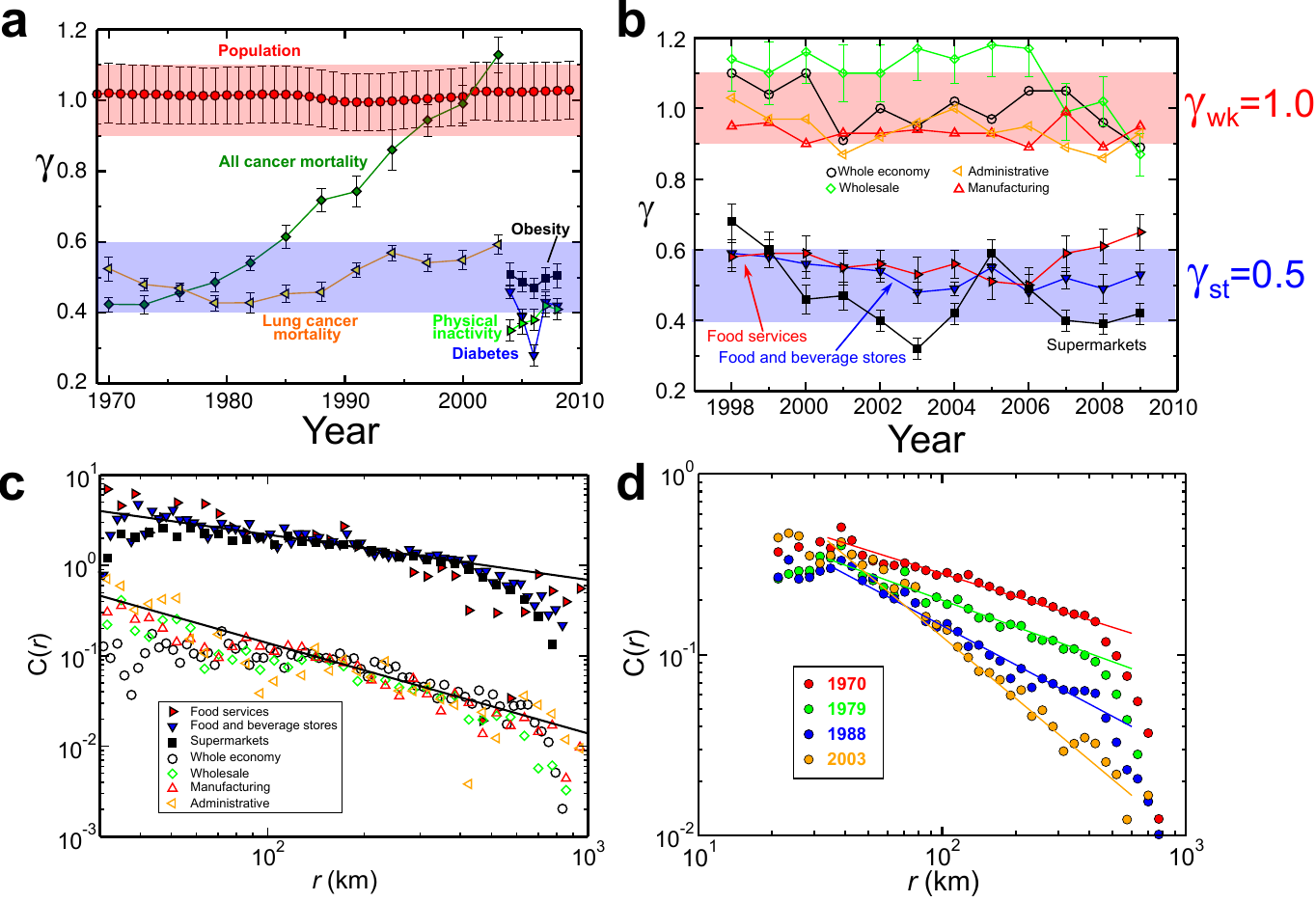}}}
\caption{ {\bf Correlation exponents.}  {\bf a,} Temporal
evolution of $\gamma$ for population distribution, obesity, diabetes,
physical inactivity, all cancer mortality, and lung cancer mortality
per county.  The diagram displays the classes of strong correlations,
$\gamma_{\rm st}= 0.5$, and weak correlations, $\gamma_{\rm wk}=1$.
Additionally, theory predicts $\gamma_{\rm rnd} \ge 2$ for
uncorrelated systems.  We did not observe any human activity or
indicators whose correlations fall within this class, unless the data
of different counties is shuffled. {\bf b,} Evolution of $\gamma$ for
different economic indicators describing the food industry, the whole
economy and generic economic sectors as indicated.  We quantify
economic activity by the total number of employees of a given sector
per county population. {\bf c,} Correlation functions for the
economic activities indicated in the figure. The plot shows the
segregation of the data into two classes.  The solid lines indicate
$\gamma_{\rm wk} = 1$ and $\gamma_{\rm st}=1/2$.  {\bf d,} Change in
$C(r)$ for cancer mortality rates in the period 1970-2003.
}
\label{time}
\end{figure}

We interpret any departure from $\gamma=1$ as a proxy of anomalous
dynamics beyond the simple dynamics related to the population growth.
When we calculate the spatial correlations of obesity prevalence
($s_i\equiv o_i/p_i$, $o_i$ is the number of obese adults in county
$i$) in USA from 2004 to 2008 we also find long-range correlations
(Fig. \ref{correl}a).  The crux of the matter is that the correlation
exponent for obesity ($\gamma=0.50\pm0.04$, averaged over 2004-2008)
is smaller than that of the population, signaling anomalous
growth. Since smaller exponents mean stronger correlations, the
increase in obesity prevalence in a given place can eventually spread
significantly further than expected from the population dynamics.

We also calculate fluctuations in variables which are known to be
strongly related to obesity
\cite{popkin,cutler,schuurman,Mokdad_prevalence_of_obesity}: diabetes
and physical inactivity prevalence (fraction of adults per county who
report no physical activity or exercise, see Methods Section \ref{data}).
The obtained $\gamma$ exponents are anomalous with similar values
as in obesity (Fig. \ref{time}a).
The system size dependence of $\xi$ for obesity and diabetes cannot be
measured directly, since there is no available micro-data for other
countries. However, we find that $\xi$ of obesity and diabetes in USA
is very close to $\xi$ of the population distribution (inset of
Fig. \ref{correl}a). Assuming that the equality of the correlation
lengths holds also for other countries, then obesity and diabetes
should satisfy Eq. (\ref{sf}) as well.




The correlations in obesity are reminiscent of those in physical
systems at a critical point of second-order phase transitions
\cite{stanley,mora,cavagna}. Physical systems away from criticality
are uncorrelated and fluctuations in observables, e.g., magnetization
in a ferromagnet or density in a fluid, decay faster than a power-law,
e.g., exponentially \cite{stanley,mora}.
Instead, long-range correlations appear at critical points of phase
transitions where fluctuations are not independent and, as a
consequence, fall-off more slowly. The existence of long-range
correlations with $\gamma=0.5$--- rather than the noncritical
exponential decay--- signals the emergence of strong critical
fluctuations in obesity and diabetes spreading.  The notion of
criticality, initially developed for equilibrium systems
\cite{stanley,mora}, has been successfully extended to explain a wide
variety of dynamics away from equilibrium ranging from collective
behaviour of bird cohorts, biological and social systems to city
growth, just to name a few \cite{cavagna,mora,rozenfeld,stanley3}. Its
most important consequence is that it characterizes a system for which
local details of interactions have a negligible influence in the
global dynamics \cite{stanley,mora}. Following this framework, the
clustering patterns of obesity are interpreted as the result of
collective behaviors which are not merely the consequence of
fluctuations of individual habits.


As a tentative way of addressing which elements of the economy may be
related to  the obesity spread, 
we calculate $\gamma$ in economic indicators which are thought to be
involved in the rise in obesity \cite{swinburn,nestle}.  Except for
transient phenomena, all studied indicators yield exponents that fall
around $\gamma_{\rm wk} = 1$ or $\gamma_{\rm st} = 1/2$, representing
two universality classes of weak and strong correlations, respectively
(Figs. \ref{time}a and b).

We begin by studying the correlations in the whole economy (measured
through the number of employees of all economic sectors per county
population, see Methods Section \ref{data}). We find $\gamma$ close to
$\gamma_{\rm wk}=1$ (over the period 1998-2009, Fig. \ref{time}b and
c) suggesting that the whole economy inherits the correlations in the
population. Generic sectors of the economy which are not believed to
be drivers of obesity, e.g., wholesalers, administration, and
manufacturing, also display $\gamma$ consistent with the population
trend (Fig. \ref{time}b and c).

Interestingly, analysis of the spatial fluctuations in the economic
activity of sectors associated to food production and sales
(supermarkets, food and beverages stores and food services such as
restaurants and bars) gives rise to the same anomalous value as
obesity and diabetes ($\gamma_{\rm st}= 1/2$, 1998-2009,
Fig. \ref{time}b and c).  Although these results cannot inform about
the causality of these relations,
they show that the scaling properties of the obesity patterns display
a spatial coupling which is also expressed by the fluctuations of
sectors of the economy related to food production.  

\begin{figure}
\centerline{\resizebox{8.0cm}{!} { \includegraphics{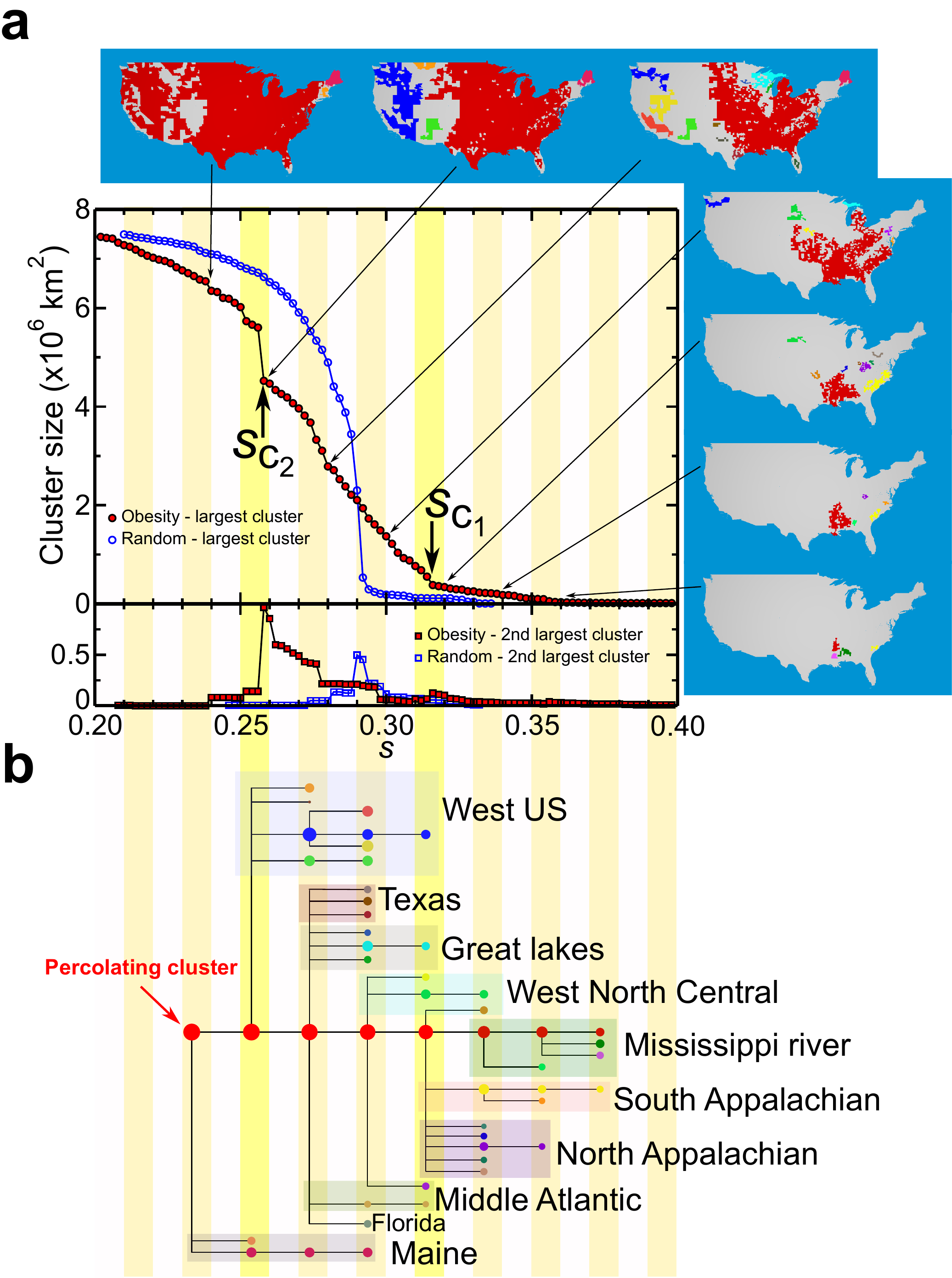}}}
\caption{{\bf Percolation picture of obesity.}  {\bf
  a,} Size of the largest component (circles) and second largest
component (squares) as a function of the obesity prevalence threshold
$s$ in 2008.  As we lower $s$, the size of the largest component
increases by absorbing smaller clusters. In many cases, this is done
abruptly, indicating that whole clusters have been incorporated to the
largest cluster as evidenced by the peaks in the second largest
cluster and the jumps in the size of the largest clusters
\cite{shlomo}. We observe two main transitions at $s_{c_1}$ and
$s_{c_2}$ in the real data (red) and a single percolation second-order
transition in the randomized data (blue).  The maps show the
progression of the obesity clusters with at least 5 counties for a
given $s$. {\bf b,} Percolation tree representing the hierarchical
formation, growth and merging of obesity clusters. Each dot represents
a cluster at a given $s$ with a size proportional to the logarithm of
the cluster's area. Cluster colors follow Fig. \ref{percolation}a and
we separate them by the indicated geographic regions. As we lower $s$
from right to left, regions of high obesity prevalence appear first in
the tree.  We notice the main percolating cluster starting in the
lower Mississippi basin (red) at high $s$ and absorbing the other
clusters until percolating through all US. In particular, we note the
two main transitions at $s_{c_1}$, where it absorbs the two
Appalachian clusters, and at $s_{c_2}$, where it absorbs the West US cluster.
}
\label{percolation}
\end{figure}

\setcounter{figure}{3}
\begin{figure}
\centerline{\resizebox{5.0cm}{!} { \includegraphics{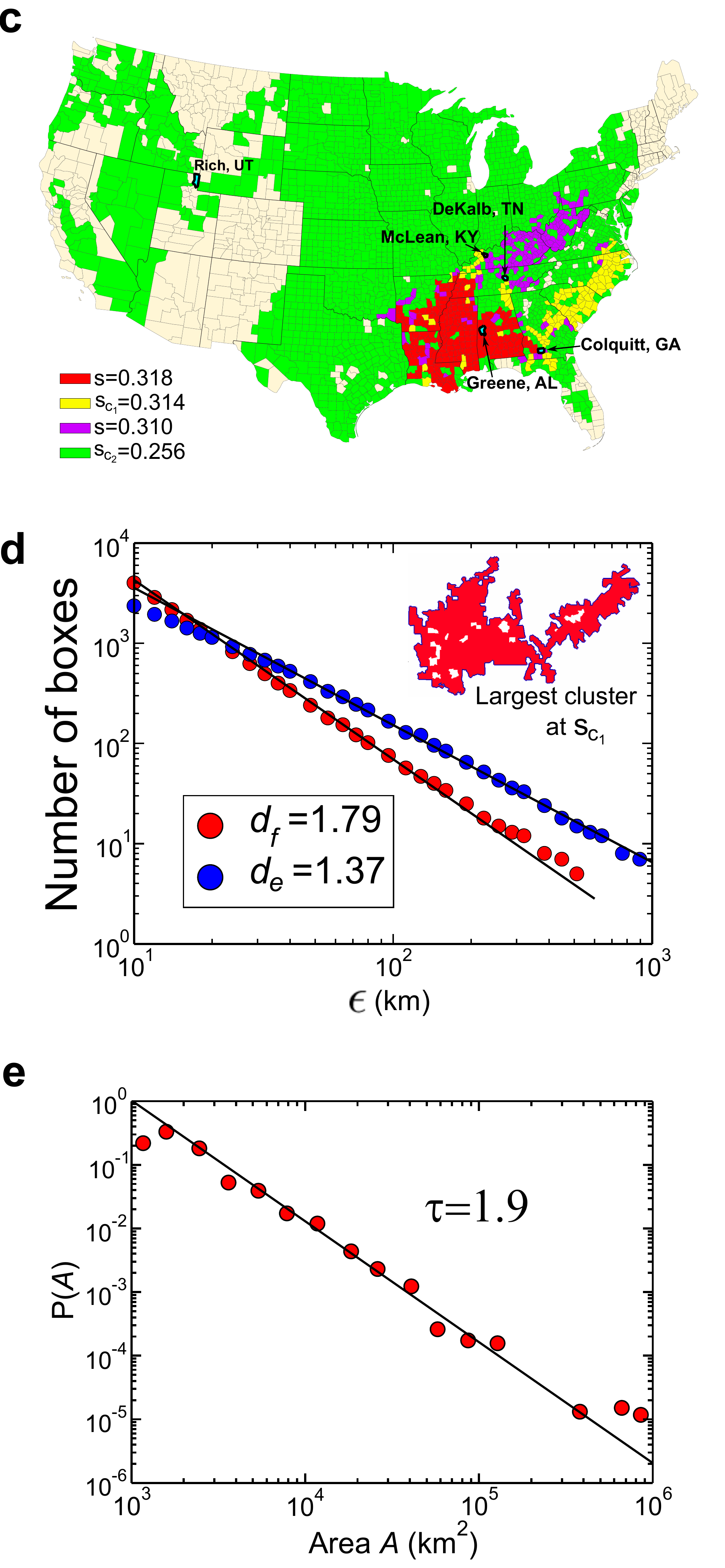}}}
\caption{{\bf c,} Detail of the evolution of obesity clusters near
  percolation as indicated.  The map shows the shape of the first
  (red), second (yellow), and third (violet) clusters around
  $s_{c_1}$, and the largest (green) cluster at $s_{c_2}$, together
  with the location of the red bonds responsible for the
  transitions. The epicenter is Greene county, AL with 43.7\% obesity
  prevalence. {\bf d,} Box fractal dimension of percolating cluster in
  the inset measured by the number of boxes of size $\epsilon$ needed
  to cover the cluster: $N_B(\epsilon)\sim \epsilon^{-d_f}$, and
  fractal dimension of the boundary measured by the number of boxes
  needed to cover the hull: $N_h(\epsilon)\sim \epsilon^{-d_e}$. {\bf
    e,} Probability distribution of the area of the obesity clusters,
  $P(A)\sim A^{-\tau}$, at percolation $s_{c_1}$ average from
  2004-2008.  This scaling law generalizes Zipf's law \cite{rozenfeld}
  from urban to obesity clusters.  }
\end{figure}

It is of interest to study other health indicators for which active
health policies have been devoted to control the rate of growth. We
apply the correlation analysis to lung cancer mortality defined at the
county level and compare with cancer mortality due to all types (Methods Section \ref{data}).  The spatial correlations of cancer mortality per
county show an interesting transition in the late 70's from anomalous
strong correlations, $\gamma_{\rm st}=1/2$, to weak correlations,
$\gamma_{\rm wk}=1$, (Fig. \ref{time}a and d). This transition is
visualized in the different correlated patterns of cancer mortality in
1970 and 2003 in Fig. \ref{transition}d, i.e., the clustering of the
data is more profound in 1970, while in 2003 it spreads more
uniformly.
The current status of all-cancer mortality fluctuations is close to
the natural one, inflicted by population correlation.  Conversely,
fluctuations in the mortality rate due to lung cancer from 1970 to
2003 have remained highly correlated and close to the obesity value,
$\gamma_{\rm st}=1/2$ (Fig. \ref{time}a and \ref{transition}e), while
the other types of cancer have become less correlated. This is an
interesting finding since, similarly to obesity, lung cancer
prevalence is affected by a global factor (smoking) and has been
growing rapidly during the studied period.

The most visible characteristic of correlations is the formation of
spatial clusters of obesity prevalence. To quantitatively determine the
geographical formation of obesity clusters, we implement a percolation
analysis \cite{shlomo,coniglio,weinrib1,sona,makse,weinrib2}.  The
control parameter of the analysis is the obesity threshold, $s$.  An
obesity cluster is a maximally connected set of counties for which
$s_i$ exceeds a given threshold $s$: $s_i \ge s$.  By decreasing $s$,
we monitor the progressive formation, growth and merging of obesity
clusters.

In random uncorrelated percolation \cite{shlomo}, small clusters would
be formed in a spatially uniform way until a critical value, $s_c$, is
reached, and an incipient cluster spans the entire system. Instead,
when we analyze the obesity clusters we observe a more complex pattern
exemplified in Fig. \ref{percolation}a and \ref{percolation}b for year
2008.  At large $s$, the first cluster appears in the lower
Mississippi basin (red in Fig. \ref{percolation}a) with epicenter in
Greene county, AL.
Upon decreasing $s$ to 0.32, new clusters are born including two
spanning the South and North of the Appalachian Mountains, which acts
as a geographical barrier separating the second and third largest
clusters (yellow and violet in Fig. \ref{percolation}a,
respectively). Further lowering $s$, we observe a percolation
transition in which the Appalachian clusters merge with the
Mississippi cluster. This point is revealed by a jump in the size of
the largest component and a peak in the second largest component at
$s_{c_1}=0.314$ (Fig. \ref{percolation}a) as features of a percolation
transition \cite{shlomo}.  At $s_{c_1}$, three ``red bonds'' (McLean
county, KY, DeKalb county, TN, and Colquitt county, GA) appear to
connect the incipient largest cluster spanning the East of USA (see
Fig. \ref{percolation}c and Methods Section \ref{red}). As a comparison,
when we randomize the obesity data by shuffling the values between
counties, a single $s_c=0.29$ appears as a signature of a uncorrelated
percolation process (blue symbols in Fig. \ref{percolation}a).


Obesity clusters in the West persist segregated from the main Eastern
cluster avoiding a full-country percolation due to low-prevalence
areas around Colorado state. Finally, the East and West clusters merge
at $s_{c_2}=0.256$ by a red bond (Rich county, Utah) producing a
second percolation transition; this time spanning the whole country
(see Fig. \ref{percolation}a and c, where the whole spanning cluster is
green, and Methods Section \ref{red}). This cluster-merging process is a
hierarchical percolation progression represented in the tree model in
Fig. \ref{percolation}b.

To further inquire whether the spreading of obesity has the features
of a physical system at the critical point, we examine
the geometry and distribution of obesity clusters. For long-range
correlated critical systems percolating through nearest neighbors in
two dimensional maps, the
geometrical structure \cite{shlomo,sona,makse} gives rise to three
critical exponents: the fractal dimension of the spanning cluster,
$d_f$, the fractal dimension of the hull, $d_e$, and the cluster size
distribution exponent, $\tau$, analogous to Zipf's law
\cite{rozenfeld} (Methods Section \ref{som-percolation}).  For the
percolating obesity cluster at $s_{c_1}$ displayed in the inset of
Fig. \ref{percolation}d, we confirm critical scaling with exponents:
$(d_f, d_e, \tau) = ( 1.79 \pm 0.08, 1.37 \pm 0.06, 1.9 \pm 0.1)$
(Fig. \ref{percolation}d, e).

Taken together, these results show that obesity spreading behaves as a
self-similar strongly-correlated critical system
\cite{stanley,mora}. In particular, a note of caution has to be raised
since, even if the highest prevalence of obesity is localized to the
South and Appalachia, the scaling analysis indicates that the obesity
problem is the same (self-similar) across all USA, including the lower
prevalence areas.

Our results cannot establish a causal relation between obesity
prevalence and economic indicators: whether fluctuations in the food
economy may impact obesity or, instead, whether the food industry
reacts to obesity demands. However, the comparative similarities of
statistical properties of demographic and economical variables serves
to identify possible candidates which shape the
epidemic. Specifically, the observation of a common universality class
in the fluctuations of obesity prevalence and economic activity of
supermarkets, food stores and food services--- which cluster in a
different universality class than simple population dynamics--- is in
line with studies proposing that an important component of the rise of
obesity is linked to the obesogenic environment \cite{hill,obesogenic}
regulated by food market economies \cite{cutler,swinburn,nestle}.
This result is consistent with recent research that relates obesity
with residential proximity to fast-food stores and restaurants
\cite{christakis2,christakis3}.  The present analysis based on
clustering and critical fluctuations is a supplement to studies of
association between people's BMI and food's environment based on
covariance \cite{christakis2,christakis3} (Methods Section
\ref{covariance}). We have detected potential candidates in the
economy which relate to the spreading of obesity by showing the same
universal fluctuation properties. These tentative relations ought to
be corroborated by future intervention studies.

\noindent{\bf Acknowledgements:} 
LKG and HAM are supported by NSF-0827508 Emerging Frontiers,
and PB and MS by Human Frontiers Science Program. We
are grateful to H. Nickerson for bringing our attention to this
problem, and S. Alarc\'on-D\'iaz and D. Rybski for valuable
discussions. We thank Epiwork, ARL and the Israel Science Foundation
for support.



\clearpage

\centerline{\bf Methods}

\section{Datasets}
\label{data}

Obesity is determined through the Body Mass Index (BMI) which compares
the weight and height of an individual via the formula
weight(kg)/height(m$^2$).  A BMI value of 30 is considered the obesity
threshold. Overweight but not obese is $25<$BMI$<30$, and underweight
is BMI$<$18.5. Our main measure in this work is the adult obesity
prevalence of a county, $s_i=o_i/p_i$, for a given year defined as the
number of obese adults $o_i$ (BMI$>30$) in a county $i$ over the total
number of adults in this county, $p_i$.  We use the data from the USA
Center for Disease Control (CDC) downloaded from \cite{cdc}. CDC
provides an estimate of the obesity country-wide since 1970, at the
state level from 1984 to 2009, and at the county level from 2004 to
2008. The study of the correlation function $C(r)$ requires high
resolution data. Therefore, we use data defined at the county level
and restrict our study of obesity and diabetes to the available period
2004-2008.  Other indicators are provided by different agencies at the
county level for longer periods.

The datasets analyzed in this paper were obtained from the websites as
indicated below.  They can be downloaded as a single tar datafile from
\url{jamlab.org}.  The datasets consist of a list of populations and
other indicators at specific counties in the USA at a given year. A
graphical representation of the obesity data can be seen in
Fig.~\ref{transition}b for USA from 2004 to 2008, where each point in
the maps represents a data point of obesity prevalence directly
extracted from the dataset.

The datasets that we use in our study have been collected from the
following sources:

{\bf (a) Population.--} US Census Bureau. We download a number of
datasets at the county level from
\url{http://www.census.gov/support/USACdataDownloads.html}

- For the population estimates we use the table PIN030.  For the years
1969-2000 we use data supplied by BEA (Bureau of Economic Analysis)
and for years 2000-2009 we use the file CO-EST2009-ALLDATA.csv from
http://www.census.gov/popest/counties/files/CO-EST2009-ALLDATA.csv

{\bf (b) Health indicators.--} Data downloaded from the Centers for
Disease Control and Prevention (CDC).

\url{http://apps.nccd.cdc.gov/DDT_STRS2/NationalDiabetesPrevalenceEstimates.aspx}

The center provides county estimates between the years 2004-2008 for:

- Diagnosed diabetes in adults. 

- Obesity prevalence in adults. 

- Physical inactivity in adults.


The estimates for obesity and diabetes prevalence and leisure-time
physical inactivity were derived by the CDC using data from the census
and the Behavioral Risk Factor Surveillance System (BRFSS) for 2004,
2005, 2006, 2007 and 2008.  BRFSS is an ongoing, state-based,
random-digit-dialed telephone survey of the U.S. civilian,
non-institutionalized population aged 18 years and older. The analysis
provided by the BRFSS is based on self-reported data, and estimates
are age-adjusted on the basis of the 2000 US standard
population. Full information about the methodology can be obtained at
http://www.cdc.gov/diabetes/statistics.


{\bf (c) Economic indicators.--} We download data for economic
activity through http://www.census.gov/econ/. The economic activity of
each sector is measured as the total number of employees in this
sector per county in a given year normalized by the population of the
county.  The North American Industry Classification System (NAICS)
(http://www.census.gov/epcd/naics02/naicod02.htm) assigns
hierarchically a number based on the particular economy sector.  The
NAICS is the standard used by US statistical agencies in classifying
business establishments across the US business economy.

In this study we have used the following economic sectors with their
corresponding NAICS:

\begin{itemize}

\item Whole economy. Entire output of all economic sectors combined
  including all NAICS codes.

\item 31. Manufacturing. Broad economic sector from textiles, to
  construction materials, iron, machines, etc.

\item
  42. Wholesale trade. Very broad sector including merchants
  wholesalers, motors, furniture, durable goods, etc.

\item 
56. Administrative jobs and support services.

\item 445. Food and beverage stores. Including all the food sectors,
  from supermarkets, fish, vegetables meat markets, to restaurants and
  bars and other services to the food industry.

\item 44511. Supermarkets and other grocery (except convenience)
  stores. This is a subsection of NAICS 445. 

\item 722.  Food services and drinking places. A sub-sector of NAICS
  72 which includes restaurants, cafeterias, snacks and nonalcoholic
  beverage bars, caterers, bars and drinking places (alcoholic
  beverages).

\end{itemize}



{\bf (d) Mortality rates.--} We use data from the National Cancer
Institute SEER, Surveillance Epidemiology and End Results downloaded from 
http://seer.cancer.gov/data/ 

The Institute provides mortality data from 1970 to 2003, aggregated
every three years.  We analyze the mortality of a specific form of
cancer per county normalized by the population of the county.  Here, we
use mortality data for the following causes of death:


- All cancer, independently of type.

- Lung cancer.


\subsection{Gridded data of population from CIESIN}
\label{gridded}

We take advantage of the available data of population distribution
around the globe defined in a square grid of 2.5 arc-seconds obtained
from \cite{ciesin}.  This data allows to study the correlation
functions of the population distribution for many countries.  By using
this data we are able to test the system size dependence of our
results.
We find that the correlation length $\xi$ is proportional to the
linear size of the country, $L$.  The linear size is calculated as
Total Area = $L^2$. We find that the correlation scales with the
system size as discussed in the text. For instance, for the USA
population distribution we find $\xi=1050$km, while a smaller country
like UK has $\xi=321$ km.

Table \ref{xi} shows a list of countries used in
Figs. \ref{correl}b and c to determine the correlation length $\xi$ of
the correlation function of population density.


\section{Evolution of obesity clusters near percolation}
\label{red}

The shape of the main obesity clusters and location of the red bonds
and obesity epicenter are depicted in Fig. \ref{percolation}c
overlayed with a US map showing the boundaries of states and counties.
Figure \ref{percolation}c shows the obesity clusters obtained at
$s=0.318$, $s_{c_1}=0.314$, $s=0.310$, and $s_{c_2}=0.256$, depicting
the process of percolation. At $s=0.318$, we plot the largest red
cluster which is seen in the lower Mississippi basin.  The highest
obesity prevalence is in Greene county, AL, which acts as the
epicenter of the epidemic. At $s_{c_1}$, we plot in yellow the second
largest cluster in the Atlantic region south of the Appalachian
Mountains, and at $s=0.310$ we plot the third largest cluster
(violet), which appears north of the Appalachian Mountains.  We mark
with black the three red bonds that make the Mississippi cluster to
grow abruptly by absorbing the clusters in the Appalachian range. The
red bonds are DeKalb county, TN, McLean county, KY, and Colquitt
county, GA. This transition is reflected in the jump in the size of
the largest cluster in Fig. \ref{percolation}a.  The same process is
observed in the second percolation transition at $s_{c_2}$, when the
red bond, Rich county, UT, joins the Eastern and Western clusters for
a whole-country percolation.

\section{Scaling exponents of percolation clusters}
\label{som-percolation}

The scaling properties characterizing the geometry and distribution of
clusters at percolation are \cite{shlomo}:

{\it (i)} The scaling of the number of boxes $N_B$ to cover the
infinite spanning cluster versus the size of the boxes $\epsilon$:
\begin{equation}
N_B(\epsilon) \sim \epsilon^{-d_f},
\end{equation}
defining the fractal dimension of the spanning cluster,
$d_f$. 

{\it (ii)} The number of boxes, $N_h$, of size $\epsilon$ covering the
perimeter of the infinite cluster:
\begin{equation}
N_h(\epsilon) \sim \epsilon^{-d_e},
\end{equation}
defining the hull fractal dimension, $d_e$. 

{\it (iii)} The probability distribution of the area of clusters at
percolation:
\begin{equation}
P(A) \sim A^{-\tau},
\label{tau}
\end{equation}
characterized by the critical exponent $\tau$. Additionally, there is
a scaling relation between the fractal dimension and the cluster
distribution exponent \cite{shlomo}: $\tau=1+2/d_f.$

The exponents $(d_f, d_e, \tau)$ for percolation with long-range
correlations have been calculated numerically in \cite{sona,makse} as
a function of the correlation exponent $\gamma$ using standard
percolation analysis. There exist also a theoretical prediction based
on Renormalization Group in \cite{weinrib1} for the correlation length
exponent.  The values of $(d_f, d_e, \tau)$ for the obesity clusters
at the first percolation transition, $s_{c_1}$, are reported in the
main text. 
A direct computer simulations of long-range percolation
\cite{sona,makse}
for $\gamma=0.5$ finds the values of the three geometric exponents to
be $(d_f, d_e, \tau)=(1.9\pm0.1, 1.39\pm0.03, 2.05\pm0.08)$,
consistent with those reported here.

We notice that the exponent $\tau$ is expected to be larger than
2. This is due to mass conversation, assuming that the power-law
Eq. (\ref{tau}) extends to infinity at percolation in a infinite
system size. The fact that we find a value slightly smaller than 2 for
the obesity clusters, might be due to a finite size effect. We also
notice that the values of the exponents obtained from correlated
percolation at $\gamma=0.5$ are not too far from those of uncorrelated
percolation \cite{sona}. Therefore, the values of the exponents may
not be enough to precisely compare the obesity clusters with
long-range percolation clusters. However, they serve as an indication
that the obesity clusters have the geometrical properties of clusters
at a critical point, such as scaling behavior.  Furthermore, it could
be possible that long-range correlated percolation may capture only
part of the dynamics of the clustering epidemic. It could be, for
instance, that higher order correlations, beyond the two-point
correlation captured by $C(r)$, are also relevant in determining the
value of the exponents.  In this case, our analysis should be
supplemented by studies of $n-$point correlation functions, beyond
$C(r)$.

\begin{table}[ht]
  \caption{List of countries used to calculate the correlation length,
    $\xi$, from the correlation function of the population
    density. Data is obtained from \cite{ciesin}. $L$ is calculated as
    the square root of the total area of the country.}
  \begin{center}
\begin{tabular}{|c|r|r|}
  \hline
Country name &    $L$ (km) & $\xi$ (km)       \\
  \hline
Liechtenstein & 12.65 & 11.5 \\
Malta & 17.75 & 10 \\
Andorra & 21.54 & 8.2 \\
Bahrain & 24.96 & 14.5 \\
Hong-Kong & 32.46 & 13.7 \\
Luxembourg & 50.86 & 26.24 \\
Cyprus & 96.29 & 14.5 \\
Kuwait & 131.37 & 27.5 \\
Slovenia & 142.21 & 26 \\
El Salvador & 142.40 & 33.5 \\
Burundi & 158.83 & 71.5 \\
Albania & 168.36 & 67.8 \\
Belgium & 174.79 & 98.3 \\
Switzerland & 197.42 & 85 \\
Netherlands & 203.38 & 84 \\
Dominican Republic & 219.29 & 67.7 \\
Lithuania & 254.94 & 26.5 \\
Ireland & 263.58 & 50.5 \\
Czech Republic & 280.38 & 35.2 \\
Hungary & 303.39 & 46.5 \\
Bulgaria & 333.63 & 38.5 \\
Uruguay & 417.11 & 114 \\
United Kingdom & 497.18 & 321 \\
Oman & 551.53 & 290 \\
Poland & 557.85 & 202 \\
  \hline
\end{tabular}
\end{center}
\label{xi}
\end{table}

\begin{table}
\caption{Continuation}
  \begin{center}
\begin{tabular}{|c|r|r|}
  \hline
Country name &    $L$ (km) & $\xi$ (km)       \\
  \hline
Congo & 585.86 & 440 \\
Germany & 596.68 & 123 \\
Japan & 609.68 & 157 \\
Zimbabwe & 623.74 & 81 \\
Paraguay & 629.19 & 330 \\
Iraq & 656.18 & 142 \\
France & 739.68 & 83 \\
Kenya & 761.33 & 345 \\
Ukraine & 767.08 & 82 \\
Madagascar & 770.04 & 227 \\
Zambia & 863.317 & 190 \\
Pakistan & 886.183 & 438 \\
Nigeria & 950.913 & 293 \\
Venezuela & 954.756 & 565 \\
Bolivia & 1034.09 & 350 \\
Ethiopia & 1060.05 & 370 \\
South Africa & 1103.47 & 588 \\
Iran & 1261.09 & 618 \\
Saudi Arabia & 1392.42 & 618 \\
Mexico & 1393.92 & 622 \\
Congo Democratic Republic & 1520.99 & 420 \\
India & 1791.57 & 495 \\
Australia & 2763.09 & 565 \\
Brazil & 2912.11 & 660 \\
United States of America & 3034.92 & 1050 \\
  \hline
\end{tabular}
\end{center}
\end{table}

\section{Covariance}
\label{covariance}

Our approach supplements covariance analysis
\cite{christakis2,christakis3}.  Instead, we use physics concepts to
shed a different view on the spreading of epidemics.  Our approach can
be extended to the study of the geographical spreading of any
epidemic: from diabetes and lung cancer, to the spreading of viruses or
real states bubbles, where the spatial spreading plays an important
role.

Population correlations are naturally inherited by all demographic
observables. Even variables whose incidence varies randomly from
county to county would exhibit spatial correlations in their absolute
values, simply because its number increases in more populated counties
and population locations are correlated. Indeed, the absolute number
of obese adults per county is directly proportional to the population
of the county [Bettencourt, L. M. A., {\it et al.}
Growth, innovation, scaling, and the pace of life in cities. {\it
  Proc. Natl. Acad. Sci. USA} {\bf 104}, 7301-7306 (2007)].
Our aim is to measure spatial fluctuations on the frequency of
incidence, independent of population agglomeration. Thus, spatial
correlations of all indicators ought to be calculated on the density
defined, in the case of obesity, as $s_i = o_i/p_i$ rather than on the
absolute number of obese people, $o_i$, itself. The spatial
correlations of the fluctuations of $s_i$ from the global average
captures the collective behavior expressed in the power-law described
in Eq. (\ref{gamma}).


\end{document}